\documentclass[lettersize,journal]{IEEEtran}

\usepackage{amsfonts}
\usepackage{amssymb}
\usepackage{amsthm}
\usepackage{amsmath,amsfonts,amssymb}
\usepackage[dvips]{graphicx}
\usepackage{subfig}
\usepackage{verbatim}
\usepackage{setspace}
\usepackage{bm}
\usepackage{algorithmic} 
\usepackage[ruled,vlined]{algorithm2e}
\usepackage{cite}

\usepackage{changepage}
\usepackage{pdfpages}
\usepackage{color}

\usepackage{caption}

\IEEEoverridecommandlockouts

\begin{document}
\bstctlcite{IEEEexample:BSTcontrol} 
\title{{\fontsize{23.8}{16}\selectfont Movable Antenna-Enhanced Secure Communication: Opportunities, Challenges, and Solutions}}
	
	\author{Yaodong Ma, Kai Liu, Lipeng Zhu, Yanming Liu, Yanbo Zhu, and Daniel Benevides da Costa
	\thanks{Y. Ma, K. Liu, and Y. Zhu are with the School of Electronics and Information Engineering, Beihang University, China, and the State Key Laboratory of CNS/ATM, China (email: \{yaodongma, liuk, zhuyanbo\}@buaa.edu.cn).}		
	\thanks{L. Zhu (corresponding author) is with the Department of Electrical and Computer Engineering, National University of Singapore, Singapore (email: zhulp@nus.edu.sg).}
	\thanks{Y. Liu is with the China Satellite Network System Company Ltd., China (email: liuyanming@buaa.edu.cn).}
	\thanks{D. B. da Costa is with the Interdisciplinary Research Center for Communication Systems and Sensing, Department of Electrical Engineering, King Fahd University of Petroleum \& Minerals (KFUPM), Dhahran 31261, Saudi Arabia (email: danielbcosta@ieee.org).}
	\vspace{-0.5 em}
}
\maketitle

\begin{abstract}
The broadcast nature of wireless communication renders it inherently vulnerable to security threats such as jamming and eavesdropping. While traditional array beamforming techniques help to mitigate these threats, they usually incur high hardware and processing costs, particularly in large-scale arrays with fixed-position antennas (FPAs). In contrast, movable antenna (MA) arrays can fully exploit the channel variation in spatial regions by enabling flexible antenna movement, which has emerged as a promising technology for secure communications. This article provides a magazine-type overview of MA-aided secure communications. Specifically, we first illuminate the promising application scenarios for MA-enhanced secure communication systems. Then, we examine the security advantages of MAs over conventional FPA systems, fundamentally stemming from their ability to adjust channel correlations between legitimate users, eavesdroppers, and jammers. Furthermore, we discuss important technical challenges and their potential solutions related to MA hardware architecture, channel acquisition, and antenna position optimization to realize secure transmissions. Finally, several promising directions for MA-aided secure communications are presented to inspire future research.
\end{abstract}

\section{Introduction}
The shared spectrum and broadcast nature of wireless communication systems make them inherently vulnerable to various security threats, including active jamming for interrupting transmissions and passive eavesdropping for intercepting legitimate data \cite{zou2016survey}. Although several technologies have been proposed to address these security challenges, such as cryptographic, frequency-hopping, directional modulation, and artificial noise, these mechanisms may result in additional computational overhead and resource costs for wireless devices. In this context, the antenna arrays-based secure beamforming technique has demonstrated superior performance, leveraging its spatial degrees of freedom (DoFs) to enhance both anti-eavesdropping and anti-jamming capabilities. However, to realize ultra-secure communication, extremely large-scale antenna arrays are required using hundreds or even thousands of antennas as well as their associated radio-frequency (RF) chains, which incur exorbitant hardware costs, energy consumption, and signal processing overhead \cite{zhu2025tutorial}.

To address the limitations of existing secure beamforming approaches, movable antenna (MA) (also known as fluid antenna) \cite{zhu2023movableM} has been recognized as a promising technology to enhance secure communication performance through flexible antenna movement at the transmitter (Tx) and/or receiver (Rx). Specifically, by repositioning multiple MAs, the geometry of the antenna array is altered, which in turn modifies the steering vectors (or array response vectors) in different directions.
Thus, different from conventional fixed-position antenna (FPA) systems, the joint optimization of the antenna position vector (APV) and the antenna weight vector (AWV) in an MA array facilitates more flexible beamforming, which is beneficial to mitigate information leakage and null undesired jamming. Several initial studies have demonstrated the potential of MA arrays in improving the secrecy communication rate, achieved through one-dimensional (1D) antenna movement \cite{hu2024secure} or three-dimensional (3D) flexibility in antenna positioning \cite{maYD2025movable}. However, there is still a lack of a systematic overview of application scenarios, fundamental performance gains, and key technical challenges for MA-aided secure communications, which motivates this article to fill these gaps.



\begin{figure*}[t]
	\begin{center}
		\includegraphics[width=10.7 cm]{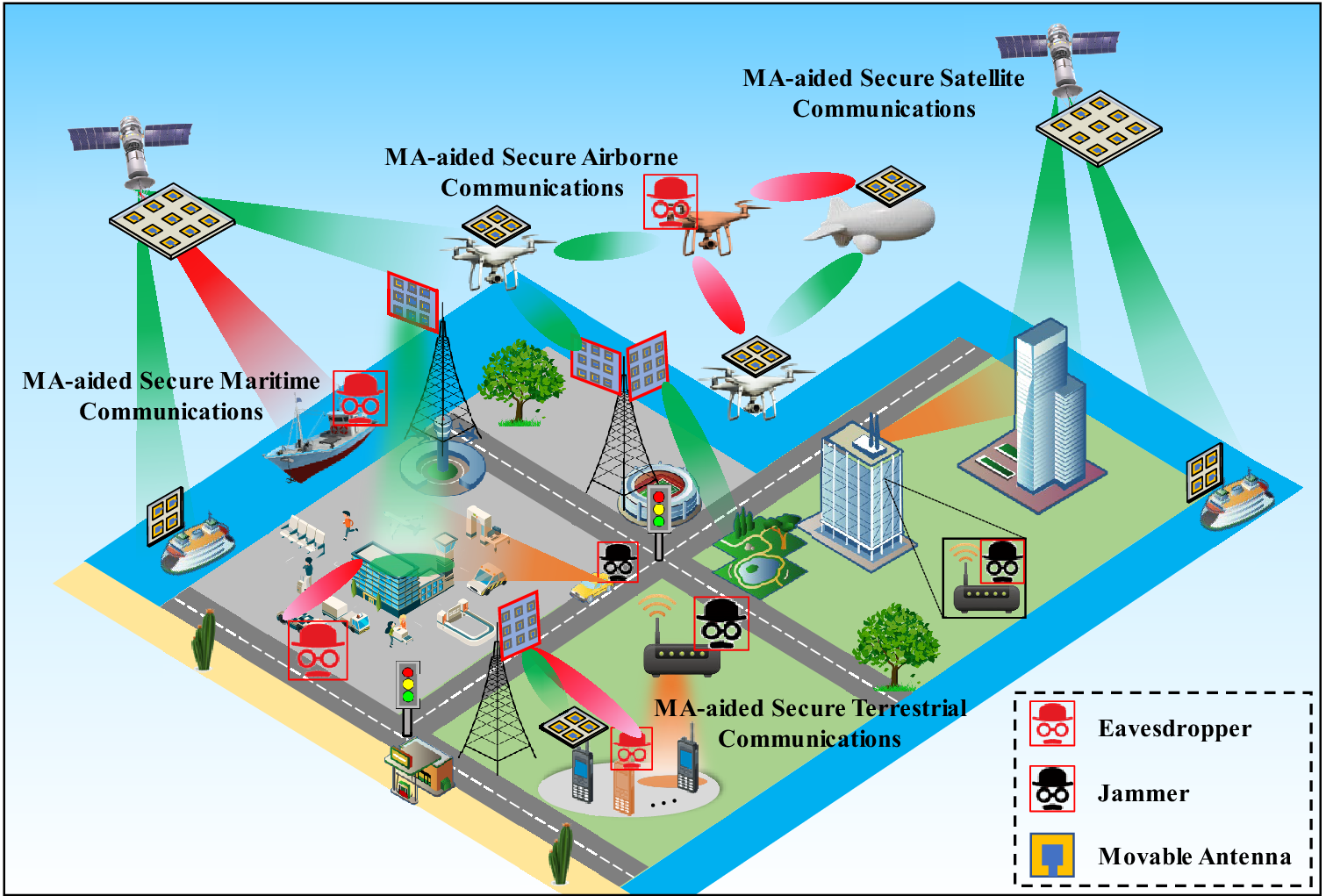}
		\caption{Typical applications for MA-aided secure communication.}
		\label{fig:MA_Applications}
	\end{center}
	\vspace{-2 em}
\end{figure*}

Figure \ref{fig:MA_Applications} presents the typical application scenarios of secure communication enhanced by MAs, including:
\begin{itemize}
	\item \emph{Terrestrial Networks}: 
	Terrestrial users, including semi-static devices (e.g., machine-type communication devices) and mobile devices (e.g., mobile phones or vehicles), are vulnerable to the attack of adversaries such as eavesdroppers and jammers. Furthermore, the existence of multiple non-line-of-sight (NLoS) propagation paths caused by abundant terrestrial scatterers increases the risk of information leakage and jamming blockage. The integration of MAs into terrestrial transceivers introduces new DoFs by enabling dynamic adjustment of antenna positions. Specifically, MA arrays can be reconfigured to reduce channel correlation between legitimate Tx/Rx and adversaries, thereby enhancing security performance, even with a limited number of antennas.
	
	
	\item \emph{Airborne Communications}: Compared to terrestrial networks, airborne communication systems can usually establish strong line-of-sight (LoS) transmission links, which not only enhance legitimate transmissions but also lead to security threats when the adversaries are closely spaced to legitimate transceivers. In such scenarios, mounting MAs on airborne platforms can enable both wavelength-scale movement and large-scale spatial displacement, offering synergetic performance gains in secure communications. 
	Specifically, the MA-mounted airborne platform can be deployed in positions with strong LoS conditions to legitimate Rx while maintaining blocked LoS conditions to adversaries. Then, by further adjusting the MA positions within wavelength-scale regions, channel correlation between legitimate Rx and adversaries can be reduced, thereby suppressing undesired signal leakage and mitigating jamming power.
			
	\item \emph{Satellite Communications}: Due to the expansive coverage areas provided by satellite communication systems, they are inherently exposed to various security risks. In particular, adversaries within the satellite's coverage are challenging to detect or neutralize, as they can conceal themselves across vast and intricate terrains, leading to significant vulnerabilities. 
	By deploying MA arrays on satellite and/or terrestrial terminals, more flexible beam shapes can be synthesized. This flexibility is achieved by jointly optimizing the spatial positions and beamforming coefficients of the MA elements, thus improving the resolution of beams directed towards the legitimate Rx. 
			
	\item \emph{Space-Air-Ground-Sea Integrated Networks (SAGSINs)}: As a four-layer heterogeneous network with open links and dynamic topologies, SAGSINs face increasing security challenges due to cross-layer attacks. Specifically, jamming and eavesdropping attacks targeting terrestrial devices or airborne platforms can propagate through the uplink, reaching satellite communication systems \cite{guo2021survey}. This propagation may cause satellites to receive false commands, potentially leading to interference or even disruption within the SAGSINs. Deploying MA arrays on each layer of SAGSINs can help reduce the received jamming power and mitigate eavesdropping through flexible beamforming, thereby significantly enhancing the overall security of SAGSINs. 
\end{itemize}

Given the vast potential and emerging applications of MAs in enhancing secure wireless communication, this article aims to provide a magazine-type overview of the unique benefits brought by MA-aided solutions, while also identifying the primary challenges in their practical implementation. In the following sections, we first reveal the fundamental gains of security performance offered by the integration of MAs. Then, we discuss the technical challenges involved in implementing MA-aided secure communication systems, and introduce potential solutions to tackle them. Finally, several promising research directions are highlighted to inspire future research in this fertile area.

\section{Fundamental Security Performance Gains}
In this section, we first reveal the fundamental advantages of the MA array compared to the FPA array in secure communications from the perspective of channel correlation reconfiguration. Then, we demonstrate the flexible beam patterns of MA arrays for enhancing secure communications in both the far-field and near-field regions.

\subsection{Channel Correlation Reconfiguration}

\subsubsection{Reduce Channel Correlation Between Legitimate Transceivers and Adversaries}
\begin{figure*}[t]
	\begin{center}
		\includegraphics[width=11 cm]{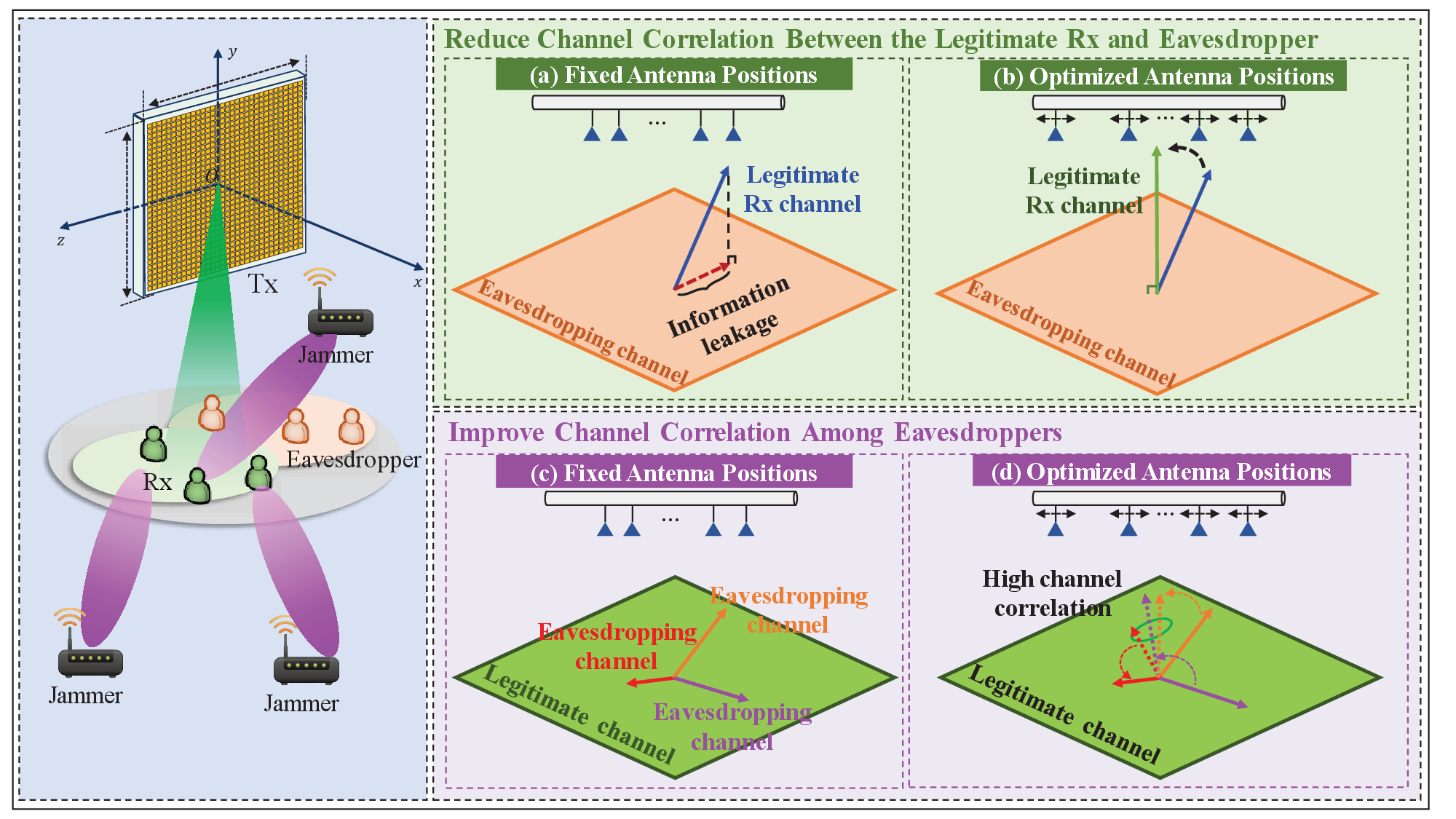}
		\caption{Illustration of the fundamental security performance gains achieved by MA arrays.}
		\label{fig:Illustration_Gain}
	\end{center}
	\vspace{-2 em}
\end{figure*}
In conventional FPA-based systems, only the antenna/beamforming weights can be reconfigured to distinguish between the legitimate Rx and the eavesdropper, which limits the flexibility in beamforming, particularly when their directions are closely aligned. In contrast, the flexible geometry of the MA array can further adjust the correlation between steering vectors associated with the legitimate Rx and the eavesdropper. As shown in Figure~\ref{fig:Illustration_Gain}, optimizing antenna placement at the Tx side can effectively decorrelate steering vectors linked to the desired user and the eavesdropping direction, yielding nearly orthogonal array responses. This facilitates the isolation of legitimate signals into distinct eigenchannels, thereby reducing leakage toward potential eavesdroppers.
Moreover, due to the inherent correlation between the legitimate and eavesdropping channels, transmit beamforming inevitably involves a trade-off between enhancing the desired signal and suppressing information leakage. 
As such, it is desirable to decrease this channel correlation, and the Tx-side MA array offers the capability to reduce the correlation between legitimate and eavesdropping channels, thereby enhancing the secrecy performance, even when the legitimate Rx is located in a direction close to the eavesdropper.
Similarly, to effectively mitigate jamming interference, the geometry of the Rx-side MA array can also be dynamically adjusted.

\subsubsection{Improve Channel Correlation Among Adversaries}
As demonstrated in Figure \ref{fig:Illustration_Gain}, we consider the scenario where threats arise from multiple eavesdroppers. In conventional FPA schemes, although transmitters are capable of suppressing signal reception at the eavesdroppers via beamforming, this inevitably results in a degradation of the desired signal power. This trade-off becomes increasingly severe as the number of eavesdroppers grows, since the beamforming flexibility has to be sacrificed to nullify the eavesdropping links, thereby depleting the available spatial DoFs for signal transmission. In contrast, the MA array is capable of aggregating the steering vectors of eavesdroppers into compact subspaces. Specifically, by strategically repositioning the MA arrays at the Tx, the eavesdroppers exhibit a high channel correlation within a low-rank channel subspace. This enables the legitimate beam to be directed as orthogonally as possible to the eavesdropping channel space, thereby minimizing information leakage. Similarly, the anti-jamming capability can be enhanced by appropriately adjusting the Rx-side antenna position of the MA array.


\begin{figure}[t]
	\begin{center}
		\includegraphics[width=8 cm]{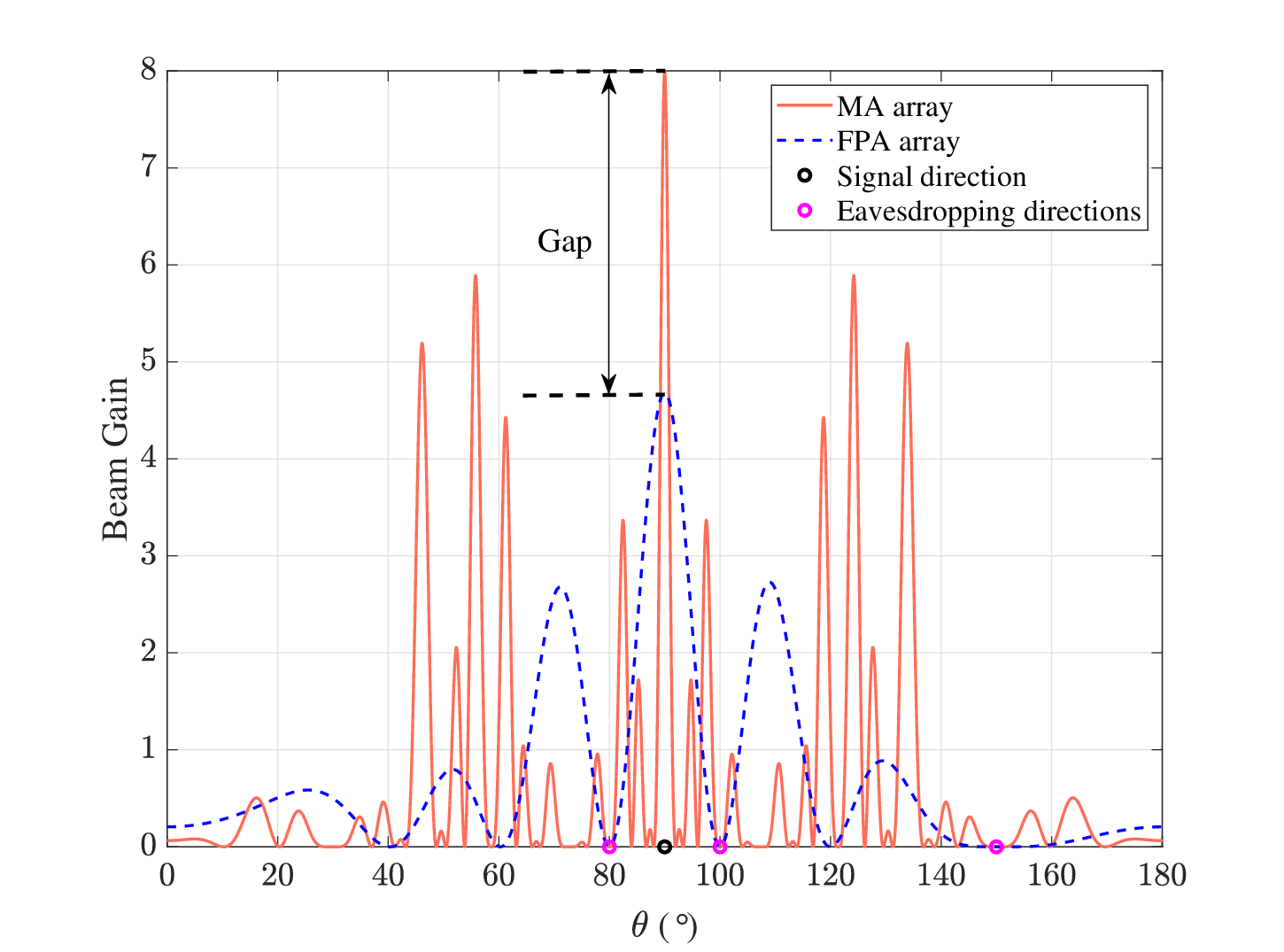}
		\caption{Comparison of beam patterns for secure communications using FPA and MA arrays.}
		\label{fig:Gain}
	\end{center}
		\vspace{-2 em}
\end{figure}

To evaluate the above channel correlation reconfiguration gains provided by MA arrays, we compare the beam patterns between MA and FPA arrays under a representative eavesdropping scenario, as illustrated in Figure \ref{fig:Gain}. Specifically, we assume a single FPA at both the Tx and eavesdroppers. Besides, both the Tx-side MA and FPA array configurations utilize 8 antenna elements, with the FPA scheme employing a uniform linear array (ULA) spaced at half a wavelength ($\lambda/2$), where $\lambda$ denotes the wavelength. The MA elements are assumed to move along to 1D line of length $A = 10\lambda$. The direction of the desired signal is denoted as $\theta_0$, and $K=3$ undesired eavesdropping angles are given by the set $\{\theta_k\}_{k=1}^{K}$, with values $\theta_0 = 90^\circ$, $\theta_1 = 80^\circ$, $\theta_2 = 100^\circ$, and $\theta_3 = 150^\circ$. The FPA-ULA scheme utilizes zero-forcing beamforming, while the antenna positions of the MA array are optimized by the beam nulling algorithm proposed in \cite{zhu2025tutorial}. Based on the above setup, as shown in Figure \ref{fig:Gain}, the MA array simultaneously achieves the full array gain  in the signal direction and precise nulls in all eavesdropping directions. Additionally, the MA array's beam exhibits significantly sharper main lobes towards the legitimate signal direction. In contrast, the FPA array experiences a loss of around $58.5\%$ in beamforming gain over $\theta_0$, resulting in a degradation of secure communication performance. This result highlights that the optimized position of the MA array contributes to enhanced angular resolution, thereby improving security performance through stronger directional beams.

\subsection{Enhancement of Near-Field  Beam Focusing}
\begin{figure*}[t!]
	\begin{center}
		\includegraphics[width=15 cm]{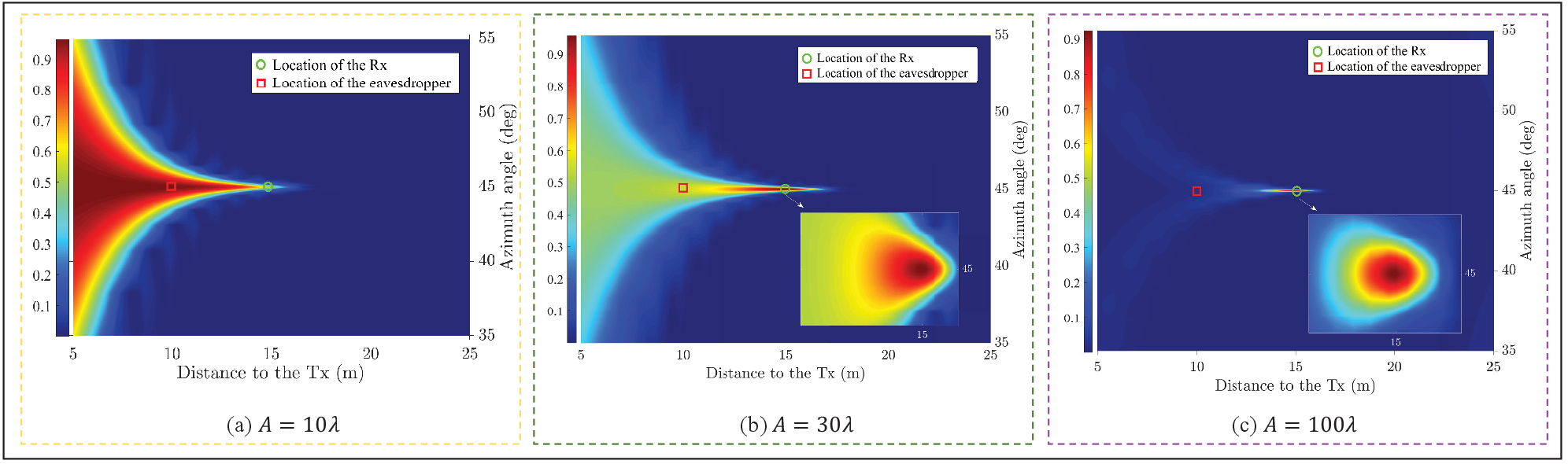}\\
		\caption{Normalized heat-maps for beam focusing with different sizes of the antenna moving region.}
		\label{fig:beamfocusing}
	\end{center}
	\vspace{-2 em}
\end{figure*}

Extremely large-scale antenna array (ELAA) has emerged as a key enabling technology for next-generation wireless networks, offering substantially improved beamforming gains. This improvement is primarily attributed to the significantly enlarged aperture size of ELAAs.
This trend, combined with higher carrier frequency bands (e.g., millimeter-wave (mmWave) and terahertz (THz)) in the next-generation networks, renders the far-field plane-wave channel model inaccurate and necessitates the use of the near-field spherical-wave model. 
Furthermore, the near-field beam focusing capability of ELAAs allows the transmitted energy to be concentrated in both angular and distance domains, thereby enhancing the power of desired communication signals, suppressing information leakage, and improving transmission security.
However, the implementation of ELAA requires the deployment of hundreds or even thousands of antenna elements, phase shifters, and their corresponding RF chains. For instance, in a square region of size 100 wavelengths, an array with half-wavelength inter-element spacing would consist of a total of $4 \times 10^4$ antennas.
In contrast, MA-aided secure communication systems can typically achieve larger aperture sizes compared to their conventional FPA-based counterparts by simply enlarging the antenna moving regions \cite{liu2023near}. 
Specifically, spherical wavefronts enable beam energy to be focused precisely at the location of the legitimate Rx, rather than merely its direction, thereby improving transmission security. As such, the legitimate Rx and adversaries with small angle/distance differences can be distinguished by MA arrays.

To reveal the security performance advantages provided by MA arrays in near-field conditions, we consider a multiple-input single-output (MISO) scenario involving large-scale Tx-side MA arrays in Figure \ref{fig:beamfocusing}, where
the MAs move within a square region of size $A \times A$.
Figure \ref{fig:beamfocusing} illustrates the beam patterns for different normalized MA moving region sizes, $A/\lambda$. 
We consider that the Tx-side MA array is equipped with 64 transmit MAs, while the Rx and eavesdropper are equipped with a single FPA. 
The carrier frequency is set to 30 GHz (i.e., $\lambda$ = 0.01 m). 
Under such parameter settings, the Rayleigh distance, i.e., the limit of the near-field region, is approximately 100 meters. Besides, the Tx, Rx, and eavesdropper are positioned at the origin, (15m, $\pi/4$), and (10m, $\pi/4$) in polar coordinates, respectively. 
The optimal position for each Tx-side MA is then determined sequentially, aiming to maximize the secrecy capacity, which is defined as the capacity difference between the legitimate Rx and the eavesdropper.
As shown in Figure \ref{fig:beamfocusing}, increasing the moving region size (i.e., from $A=10\lambda$ to $A=100\lambda$) leads to a narrower main lobe of the beam due to the larger maximum aperture that the MA array can achieve. This leads to sharper beams and finer spatial resolution, which mitigates eavesdropping and enhances communication security. In other words, by focusing the beam on the intended position, information leakage to other undesired regions is minimized. Similarly, optimizing the positions of the Rx-side MA array helps reduce jamming interference and also enhances communication performance.

\begin{figure}[t]
	\begin{center}
		\includegraphics[width=8.8 cm]{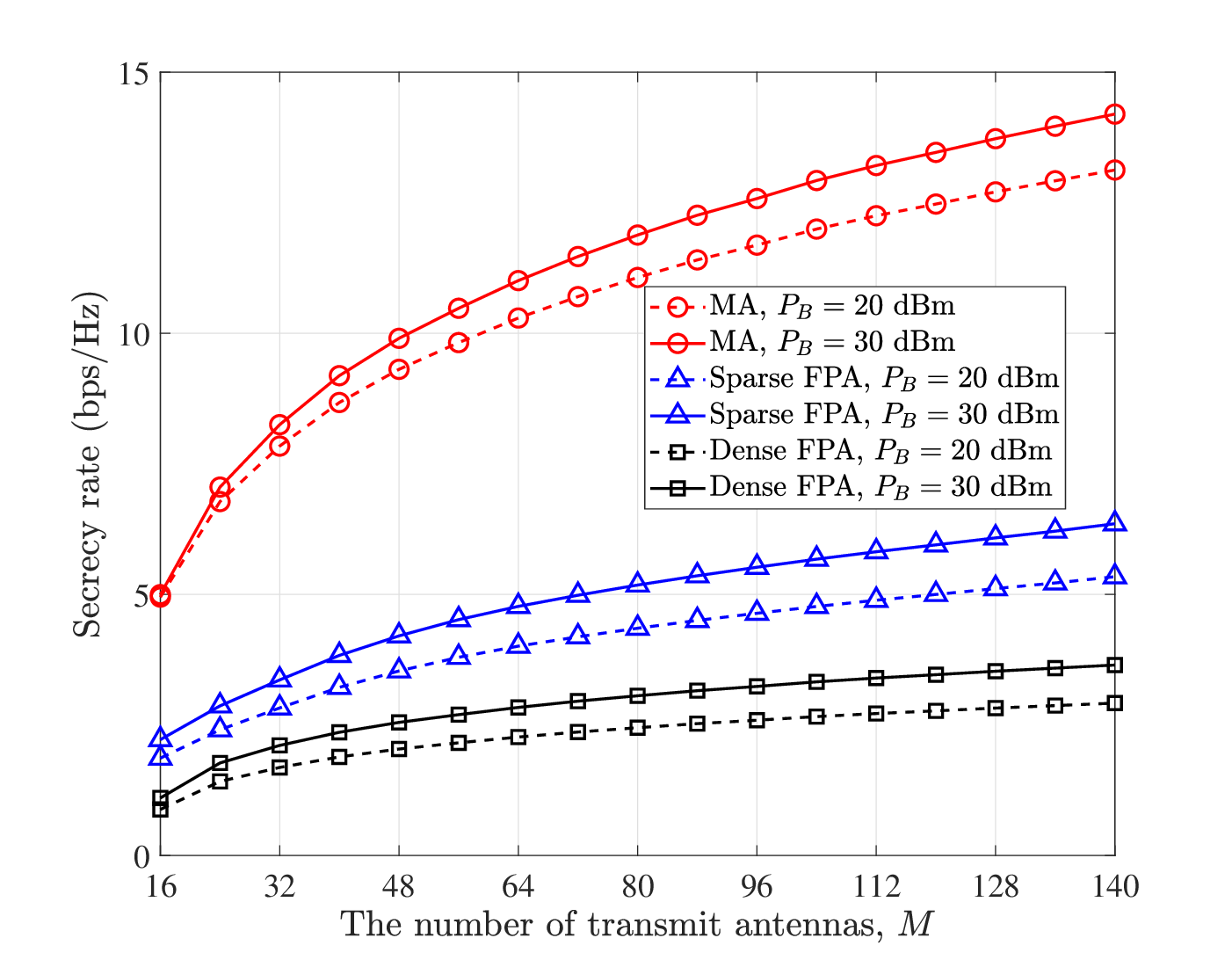}\\
		\caption{Secrecy rate versus the number of transmit antennas.}
		\label{fig:SR_vs_M}
	\end{center}
		\vspace{-2 em}
\end{figure}

In addition, Figure \ref{fig:SR_vs_M} illustrates the secrecy rate versus the number of transmit antennas at the transmitter, denoted by $M$, under various transmit power budgets, $P_B$. In addition to the MA scheme, we consider sparse FPA and dense FPA schemes as benchmarks for performance comparison. Specifically, for the dense FPA scheme, the antennas are densely arranged to form a uniform planar array (UPA) of size $8 \times 8$, with inter-antenna spacing of $\lambda/2$. In contrast, the sparse FPA scheme places the antennas sparsely to form a UPA of the same size, $8 \times 8$, with inter-antenna spacing of $A/8$. The antenna movement region is set as $A = 100\lambda$. It can be seen that as $M$ increases, the secrecy rate improves for both the MA and FPA schemes. Furthermore, the MA scheme consistently outperforms both the dense and sparse FPA schemes across all values of $M$, achieving improvements of up to 10.2 bps/Hz at $P_B = 20$ dBm and 10.6 bps/Hz at $P_B = 30$ dBm, respectively. This is because the MA system can leverage greater spatial flexibility, optimizing its beam pattern to better align with the spatial location of the legitimate Rx while nulling the beams directed towards the eavesdropper, thus achieving superior security performance. Besides, the sparse FPA scheme outperforms the dense FPA scheme as it can achieve a larger array aperture with the same number of antennas, thereby improving beam resolution between the legitimate Rx and adversaries. 
In comparison, the MA array can fully leverage both the large aperture and the geometry reconfiguration, thus achieving the best secure transmission performance.

\section{Design Issues and Solutions}
Despite the performance advantages of MA arrays in enhancing secure communications, several design issues still exist in practical secure communication systems, including implementation architecture, channel acquisition, and antenna position optimization, which are discussed in this section.

\subsection{Implementation Architectures}

\begin{figure*}[t!]
	\begin{center}
		\includegraphics[width=12.5 cm]{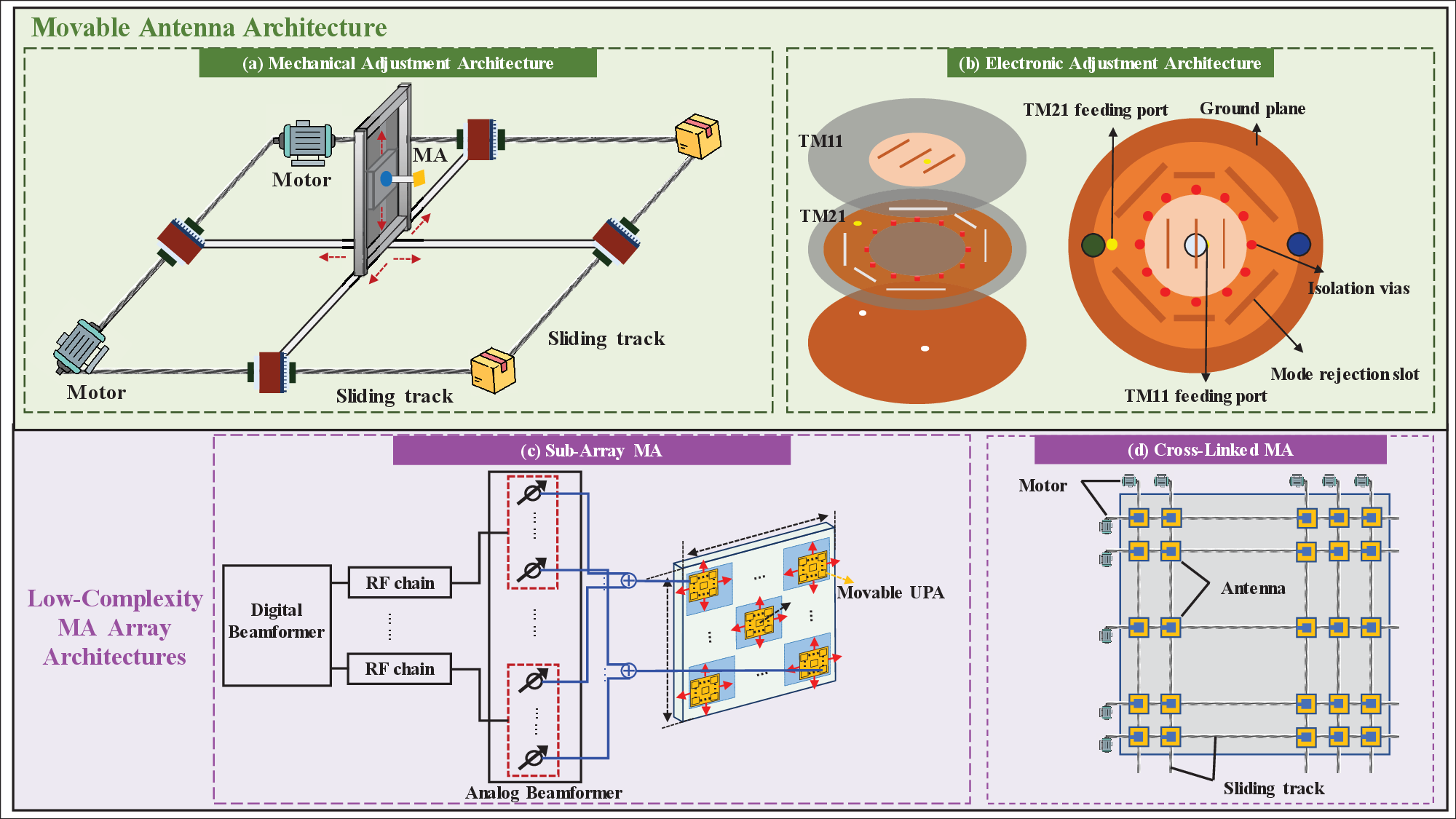}
		\caption{Architectures and hardware implementations for MA arrays.}
		\label{fig:Architecture}
	\end{center}
		\vspace{-2 em}
\end{figure*}

Figure \ref{fig:Architecture} depicts various potential hardware architectures for implementing MAs. 
Specifically, antenna positioning can be achieved using either mechanical or electronic methods, as discussed in the following sections.

\subsubsection{Mechanically Driven Movement} As shown in Figure \ref{fig:Architecture}(a), the position of each MA is adjusted by physically relocating it within a 3D space. A typical implementation employs a servo-motor-driven platform, where coordinated motor control enables the repositioning of each antenna element to desired spatial coordinates \cite{zhu2023movableM}. To further improve integration and deployment flexibility, micro-electromechanical systems (MEMS) have also been explored as an alternative actuation mechanism \cite{marnat2013new}. MEMS-based designs typically rely on torsional hinges to enable precise displacement of individual elements or subarrays, offering benefits such as compact size, fast response, and reduced power consumption. Compared to motor-driven solutions, which generally enable long-range movement, MEMS implementations are characterized by their fast response speed, and can respond at microsecond-to-millisecond timescales.
Nevertheless, the adoption of mechanically driven architectures involves trade-offs related to actuation complexity, energy efficiency, and movement latency, all of which need to be addressed to ensure antennas can quickly and accurately respond to the dynamic requirements for achieving secure communications.

\subsubsection{Electronically Driven Movement} As shown in Figure \ref{fig:Architecture}(b), electronically reconfigurable antennas, such as dual-mode patch antennas, can adjust their phase centers to emulate the displacement of active antenna positions. For example, by activating distinct resonant modes (e.g., TM11 and TM21) in a stacked circular patch, the phase center shifts from the geometric center, resulting in altered antenna positions without mechanical motion \cite{li2022using}. The electronic control architectures offer fast response times on the order of nanoseconds to milliseconds. However, practical challenges include elevated hardware costs and circuit complexity associated with achieving reconfigurability \cite{zhu2025tutorial}. Furthermore, the adjustment of antenna position is typically discrete rather than continuous, which limits the spatial resolution of beam steering and may degrade the performance of secure communications.

\subsubsection{Hybrid Architectures}
It is also feasible to integrate the aforementioned implementation architectures to achieve a more effective hybrid design tailored to specific application scenarios. For instance, in a hybrid MA array configuration, each subarray can adopt a mechanically driven architecture (as illustrated in Figure~\ref{fig:Architecture}(a)) while incorporating electronically driven antenna elements (as shown in Figure~\ref{fig:Architecture}(b)), thereby synergizing the benefits of both approaches  in terms of long-distance movement and rapid reconfiguration. Although hybrid antenna architectures provide enhanced flexibility for dynamic reconfiguration, they may lead to an increased design complexity. Therefore, careful design of low-complexity architectures is essential to fully harness their performance potential.

\subsubsection{Low-Complexity MA Array Architectures} 
The hardware cost of conventional MA systems increases with the number of movable elements, as each requires independently controllable driving components. To reduce hardware cost, more efficient MA array architectures are required. Figure \ref{fig:Architecture}(c) shows a movable subarray architecture. Instead of moving each antenna element individually, multiple subarrays are each driven by motors, allowing the antennas therein to move collectively. This approach significantly reduces implementation costs. In addition, as illustrated in Figure \ref{fig:Architecture}(d), a cross-linked MA (CL-MA) array was proposed in \cite{zhu2025multiuser}, where each MA element/subarray is mounted at the intersection of horizontal and vertical tracks. Each vertical (horizontal) track can slide along the horizontal (vertical) direction, thereby adjusting the horizontal (vertical) positions of all MAs in the corresponding column (row). As such, for an MA array with $M \times N$ elements, the conventional architecture requires $2MN$ motors with two-dimensional (2D) mobility. In contrast, the CL-MA design requires only $M+N$ motors, thereby significantly reducing hardware costs and control complexity in large-scale arrays.
In summary, developing efficient MA architectures for various secure communication systems to achieve a desired trade-off between hardware complexity and security performance is an important problem to solve in practice.

\subsection{Channel Acquisition}
The ability of the MA array to achieve secure communications heavily relies on the instantaneous or statistical channel state information (CSI), including both legitimate and jamming/eavesdropping channels. 
Since the legitimate Rx cooperates with the Tx for channel acquisition, the CSI of legitimate channels can be obtained using conventional pilot-based approaches \cite{zhu2025tutorial}. 
In general, to characterize the propagation environment, a complex-valued channel mapping from the Tx to the Rx should be constructed, capturing the responses between all spatial points in the transceiver regions. A direct measurement approach, which involves physically relocating the Tx and Rx antennas to cover every possible position, turns to be highly inefficient in large-scale systems due to significant training overhead. To address this challenge, a more scalable solution is to estimate the field-response information (FRI) in the angular domain, including angles of departure (AoDs), angles of arrival (AoAs), and multipath coefficients at both the Tx and Rx sides \cite{zhu2023modeling}. Then, by sampling the channel at a limited number of spatial positions, the angular-domain FRI can be extracted using the Fourier relationship between spatial measurements and angular components, enabling the accurate reconstruction of the full channel mapping. When the mobility of transceivers and their surrounding environment is relatively low, the positions of MAs can be dynamically adjusted to closely follow the instantaneous variations of the wireless channel. In contrast, under high-mobility scenarios with rapidly varying channels, antenna repositioning should be guided by the long-term statistical CSI, with the associated channel acquisition as an important problem for further study.

In contrast, estimating the jamming channels poses additional challenges, as adversaries do not cooperate with the transmitter for channel acquisition. As a result, the complete reconstruction of these channels is usually infeasible. Nevertheless, partial information, such as the directions or locations of jammers, can be extracted from received interference signals using beam scanning or sensing techniques \cite{schaefer2015secure}. To mitigate jamming signals, MAs at the legitimate Rx can adapt their positions and employ suitable signal processing algorithms (e.g., compressed sensing) for jammer localization. The most challenging aspect, however, is acquiring the CSI of eavesdropping channels. One possible solution is radar-assisted detection, where probing signals are transmitted, and their echoes are analyzed to recover partial CSI. In addition to instantaneous CSI estimation, statistical CSI can be inferred, enabling the approximate localization of potential eavesdroppers without requiring their precise instantaneous channels \cite{zhu2025tutorial}.

\subsection{MA Position Optimization}
The security performance gains of MA-aided secure communications over conventional FPA schemes can be attributed to their flexibility in antenna placement. However, optimizing the antenna positions to maximize security performance remains a critical challenge, since the channel response is a highly non-linear function of the MAs’ spatial locations.
As such, we discuss the MA position optimization approaches under both perfect and imperfect CSI assumptions.

\subsubsection{Perfect CSI}
In the case of perfect CSI, local optimization techniques such as gradient descent/ascent, successive convex approximation (SCA), and orthogonal matching pursuit (OMP) can be employed to obtain suboptimal placements of MAs over continuous Tx/Rx regions. When the number of MA is large, a block coordinate descent (BCD)-based approach can be applied to optimize each position sequentially. These methods offer high computational efficiency but often suffer from convergence to local optima. To address this issue, swarm intelligence-based algorithms, such as particle swarm optimization (PSO) and hippopotamus optimization, have been introduced as alternative methods.

\subsubsection{Imperfect CSI}
In the absence of perfect CSI, a straightforward approach involves performing an exhaustive search over candidate MA positions within the Tx/Rx region, selecting those that provide the best security performance. However, to ensure optimality, the number of candidate positions for MAs must be sufficiently large, which results in prohibitively high computational and energy costs, particularly in scenarios involving a large number of MAs and expansive antenna movement areas. To address this issue, additional research is required to investigate robust MA position optimization under imperfect CSI. Specifically, deterministic and stochastic models can be employed to characterize CSI uncertainties arising from estimation errors \cite{zhu2025tutorial}.
Besides, artificial intelligence (AI)-empowered methods, such as deep learning, reinforcement learning, and federated learning, can be utilized to optimize MA positions without knowing the instantaneous CSI \cite{zhu2025tutorial}. Specifically, the MA system can learn to map input features such as location, signal strength, and historical actions/positions to optimal positioning decisions through offline training, or alternatively, improve its positioning policy over time through online learning. As such, AI-based methods for designing MA-aided secure communication systems will be an interesting topic for future research.

\section{Extensions and Future Directions}
Recent advancements in antenna architectures offer promising spatial DoFs for enhancing secure communications. In this section, we investigate representative approaches, including the six-dimensional MA (6DMA), movable-element intelligent reflecting surface (IRS), and extremely large-scale MA (XL-MA) designs, each addressing unique challenges and providing innovative solutions to improve transmission security.

\subsection{6DMA-Aided Secure Communications}
To fully harness the spatial-domain gains for secure communications, the 6DMA architecture has recently been proposed \cite{shao2025tutorial}, which optimizes not only the 3D positions but also the 3D orientations of antennas, thereby introducing additional DoFs. These extra spatial DoFs enable more flexible and robust designs for secure transmission strategies. However, realizing 6DMA-assisted secure communications in practice presents several challenges. These include high-dimensional control complexity, hardware constraints in accurately actuating and sensing antenna orientations, and the need for efficient algorithms capable of jointly optimizing location and rotation parameters under real-time constraints.

\subsection{Movable-Element IRS-Aided Secure Communications}
Motivated by the flexible beamforming capabilities provided by MAs, it is promising to explore the integration of MA techniques into IRS-aided communications. Specifically, the concept of movable-element IRS envisions embedding mechanically or electronically tunable components into each IRS element, enabling precise spatial repositioning or angular adjustment of the passive elements. This architecture allows not only the traditional phase shift control but also the physical realignment of the reflective elements, thus introducing new DoFs in beam manipulation. 

\subsection{Large-Scale Antenna Movement} 
Previous studies on MAs have predominantly focused on wavelength-scale movement regions, where changes are confined to phase shifts of channel paths. In contrast, relocating the antenna over larger spatial regions, sufficient to alter the Tx–Rx large-scale fading, can establish strong LoS channels and reduce path loss for the legitimate Rx. For example, the XL-MA array proposed in \cite{fu2025extremely} facilitates flexible deployment of antenna elements or subarrays over an extremely large spatial region. By providing higher DoFs in antenna placement, the XL-MA design can significantly reduce channel correlation between the legitimate Rx and potential adversaries as well as enhance spatial resolution, which is desirable for mitigating jamming interference and rejecting eavesdropping.
Similarly, pinching antenna can also realize large-scale positioning over a pre-deployed waveguide \cite{ding2025flexible}, which has been considered for secure communication applications.

\subsection{MA-Aided Secure Sensing} 
In addition to the vulnerabilities of communication systems, radar sensing signals are also susceptible to security threats, including eavesdropping and jamming interference. MA architectures significantly enhance sensing security by dynamically adapting beam patterns, thereby improving target detection accuracy for legitimate systems while suppressing the probability of interception and jamming by unauthorized radars. In contrast, conventional FPA systems lack sufficient spatial DoFs, limiting their capability for secure sensing. Thus, MA-aided secure sensing deserves in-depth study.

\section{Conclusions}
In this article, we provided an overview of the applications, fundamental gains, challenges, and solutions for MA-aided secure communications. Essentially, MA-aided secure communication approaches offer two fundamental gains, i.e., reducing the channel correlation between the legitimate Rx and adversary and improving the channel correlation among adversaries  via antenna position optimization. 
Moreover, expanding the movement range of MAs enables sharper beams and higher spatial resolution, thereby further improving security performance. Despite these advantages, several key challenges remain, including channel acquisition, MA position optimization, and hardware architectures. Thus, more research efforts are still expected to address these challenging problems and unleash the full potential of MAs in shaping resilient and secure wireless networks.

\section*{Acknowledgments}

This work was supported in part by the National Key Research and Development Program of China under Grant 2023YFB4302800, in part by the National Natural Science Foundation of China under Grant U2133210 and Grant U2233216, and in part by the Academic Excellence Foundation of Beihang University (BUAA) for Ph.D. Students.
	
\bibliographystyle{IEEEtran} 
\bibliography{./IEEEabrv,./MAMag}

%
%
%
%
%


\end{document}